\documentclass[useAMS,usenatbib]{mn2e}

\usepackage[draft]{graphicx}

\usepackage{url,times,stfloats,amsmath,amsfonts,amssymb,aas_macros,color,epsfig,epstopdf}
\usepackage[colorlinks=true, citecolor=blue, linkcolor=blue, urlcolor=blue, pdfborder={0 0 0}]{hyperref}

\usepackage{booktabs}
\usepackage{multirow}
\usepackage{float}
\usepackage{rotating}
\usepackage{soul}

\newcommand{\hMsun}{{\ifmmode{h^{-1}{\rm {M_{\odot}}}}\else{$h^{-1}{\rm{M_{\odot}}}$}\fi}}
\newcommand{\lsim}{\lower.5ex\hbox{\ltsima}}
\newcommand{\gsim}{\lower.5ex\hbox{\gtsima}}

\newcommand{\Tab}[1]{Table~\ref{#1}}
\newcommand{\Sec}[1]{Section~\ref{#1}}

\newcommand{\Fig}[1]{Fig.~\ref{#1}}

\def\head{
 \vbox to 0pt{\vss
                   \hbox to 0pt{\hskip 440pt\rm LA-UR-10-07069\hss}
                  \vskip 25pt}}

\title[Planes of satellites in the MW and M31]
{An analysis of satellite planar configurations  around the MW and M31: singling out new high  quality planes}
\author[I. Santos-Santos et al.]
       {Isabel M. Santos-Santos$^{1,2}$\thanks{E-mail: isabelm.santos@uam.es},           
       Rosa Dom\'inguez-Tenreiro$^{1}$,
       Marcel S. Pawlowski$^{3}$\\
$^{1}$Departamento de F\'isica Te\'orica, Universidad Aut\'onoma de Madrid, 28049 Cantoblanco, Madrid, Spain \& CIAFF \\
$^{2}$Department of Physics and Astronomy, University of Victoria, Victoria, BC, Canada V8P 5C2\\
$^{3}$Leibniz-Institute f\"ur Astrophysik Potsdam, An der Sternwarte 16, D-14482 Potsdam, Germany
}

\begin{document}

\date{Accepted XXXX . Received XXXX; in original form XXXX}

\pagerange{\pageref{firstpage}--\pageref{lastpage}} \pubyear{2018}

\maketitle

\label{firstpage}


\begin{abstract}

We present a detailed characterization of planes of satellites in the Milky Way (MW) and M31 systems.  
To this end we introduce an extension to the '4-galaxy-normal density plot' method \citep{Pawlowski13}, by which
plot over-densities signal the normal direction to predominant planes of satellites within a given sample.
For a given over-density, the extension provides a \textit{collection} of planes, each including a different number of objects $N_{\rm sat}$. 
We apply this method to the position data of confirmed MW and M31 satellites and quantify the quality of planes  through the outputs of a Tensor of Inertia plane-fitting technique.
Plane quality is quantified in terms of population ($N_{\rm sat}$) and flattening (the short-to-long axis ratio $c/a$ or the rms thickness normal to the plane). Therefore, planes with the same population or flattening can be compared with each other allowing us to single-out best-quality planes.

For the first time, we study the second-most predominant planar configuration of satellites in M31, singling out a plane with 18 satellite members that shows a quality
comparable to the Great Plane of Andromeda (GPoA, with $N_{\rm sat}=19$)  despite it being more affected by distance uncertainties. This structure is viewed nearly face-on from the MW and is approximately normal to the GPoA. 

Overall, we find planes of satellites around the MW and M31 with higher qualities than those previously reported with a given  $N_{\rm sat}$. We also show that mass plays no role in determining a satellite's  membership or not to the respective best-quality planes.

\end{abstract}

\noindent
\begin{keywords}
galaxies: dwarf - Local Group - kinematics and dynamics  methods: statistical cosmology: theory
 \end{keywords}


\section{Introduction}

The known objects surrounding the Milky Way (MW) show an anisotropic spatial distribution.
\citet{Lynden76}
and \citet{Kunkel76} were
  the first to notice that 
dSphs Sculptor, Draco, Ursa Minor  and globular cluster Palomar 13, apparently lied on the orbital plane of  the Magellanic stream, therefore polar to the Galaxy.
Soon after, Fornax, Leo I, Leo II, Sextans, Phoenix
and some classified as `young halo' globular clusters, 
 were   also found to participate of this ``great circle" \citep{LyndenBell82,Majewski94,FusiPecci95}.
 The existence of a ``plane of satellites" 
in the MW
 was finally ratified when measuring the 
  very flattened distribution of the  11 classical  satellites as compared to isotropy \citep{Kroupa05,Metz07}.
In our neighboring galactic system Andromeda (M31), 
the first studies 
  by \citet{Grebel99} and \citet{Koch06}
on the then-known $\lesssim$15 dwarf galaxies within $\sim$500 kpc distance,
found 
that the subsample of dSph/dE type dwarfs 
also lied near a ``great circle". 
This  spatial anisotropy 
was emphasized by the skewness of M31 dwarfs in the direction to the MW
 \citep{McConnachie06}.
More recently,
the addition to the picture
 of newly discovered 
faint
 satellites
 thanks to surveys like SDSS \citep{York00} or PAndAS \citep{McConnachie09},
 and an increased quality of distance measurements, 
 has only enhanced the significance 
 of the planar structures noted in the MW and M31 \citep{Metz09,Kroupa10,Ibata13,Conn13,Pawlowski13,Pawlowski14,Pawlowski15b}. 
In addition,
  a richer census of young halo globular clusters and several stellar/gaseous streams 
 have been
  shown to also align with the MW satellites \citep{Keller12,Pawlowski12}. 
Finally, apart from the MW and M31, there are  claims for  a planar distribution of satellites in the nearby Centaurus A Group of galaxies \citep{Tully15}, further supported by discoveries of new dwarfs and better distance estimates \citep{Muller16,Muller18}.

 In the last years,
the quantification of 
  such    planar alignments 
  has gained increasing importance 
 in order to
 unambiguously define the observed structures
in terms of their orientation and characteristics.
These quantifications have  demanded increasingly more sophisticated methods that make use of the three-dimensional position data as well as statistical approaches to overcome measurement uncertainties or avoid spurious effects coming from working with a small sample number.

Specifically,
\citet{Koch06} used an error-weighted orthogonal distance regression accompanied by bootstrapped tests, to reliably determine a robust solution for estimating best-fit planes. 
They fitted a plane to all possible subsamples of M31 satellites 
involving 3 to 15 members
and projected the resulting normal vectors on a sphere. With the density distribution of normals they found an estimation of the normal direction to 
the  broad planar distribution defined by the satellites' positions.
\citet{Metz07,Metz09} used instead the Tensor of Inertia  plane-fitting method
taking into account distance uncertainties. 
More recently, \citet{Pawlowski13} combined the previous efforts and  presented a new statistical method to define the direction of   predominant plane-like spatial distributions  of satellites within a given sample:  the `4-galaxy-normal density plot' method, consisting of a planar fit to every combination of 4 satellites. 
 From its application  to the 
 confirmed satellites 
within 300 kpc in the MW and M31
 they obtained
the normal direction to one predominant planar alignment of satellites in each galactic system. 
In this way, 
the so-called ``VPOS-3" (Vast Polar Structure) and ``GPoA" (Great Plane of Andromeda) planes
 were defined, consisting of 24 and 19 satellites respectively.
These planes have been considered so far to be the most relevant satellite planar configurations in the MW and M31, 
and have been used as a benchmark against which to test the
 alignment of the newest (unclassified) objects discovered \citep{Pawlowski14,Pawlowski15b}. 
 However, an identification of the ``best" planes of satellites in the MW and M31  formed by a variable number of members
is still lacking. 
This identification demands an analysis of how the \textit{quality} of these planes changes with the number of satellites included.

The purpose of this paper is to present a more detailed quantification and characterization of the plane-like spatial structures in the MW and M31 satellite systems. 
This is an important issue and,
additionally, it will provide a reference with which to compare results from numerical simulations analyses (Santos-Santos in prep.). 
We will focus on the positions of satellites, which for the MW have error bars  much smaller than those of kinematic data. 
We note, however, that complementary, relevant information to satellite planar alignments comes from kinematic data.
Indeed, recent proper motions for MW satellites measured with GAIA \citep{GaiaHelmi18}
  suggest that a non-negligible fraction of them are orbiting  within the VPOS \citep{Fritz18}. 
For M31 satellites  
just line-of-sight velocities \citep{Ibata13} are currently available.

In this work we build on the results of \citet{Pawlowski13} and develop
an extension to their `4-galaxy-normal density plot' method
that enables a deeper study on 
cataloguing and quality analyses of 
planes
of satellites. 
In particular,
for each predominant planar configuration of satellites 
in the MW and M31
found with the previous method,
we yield a collection of planes of satellites  with an increasing number of members,
and identify the highest-quality
planes   in terms of the Tensor of Inertia parameters.
In particular, a plane's quality is quantified in terms of its population ($N_{\rm sat}$) and flattening (the concentration ellipsoid short-to-long axis ratio $c/a$ or, equivalently, the r.m.s. thickness normal to the plane; \citealt{Cramer}). Quality of planes with the same population or flattening can be compared with each other, allowing to single-out best quality planes with the same $N_{\rm sat}$ or $c/a$ values.

 The paper is organized as follows. 
In Sec. 2  we present the sample and dataset of MW and M31 satellites studied.
 In Sec. 3 we thoroughly describe our methodology.   Secs. 4.1 and 4.2 show the results of our quality analysis for planes in the MW and M31, respectively. Finally Sec. 5 summarizes our conclusions.

\section{MW and M31 data}\label{observedplanes}

In this study  we use the same satellite samples as in \citet{Pawlowski13}, consisting of 27 and 34 satellites for the MW and M31, respectively.
These are the confirmed satellites  within 300 kpc from their hosts, according to the 
 \citet{McConnachie12} ``Nearby dwarf galaxy database"\footnote{\url{http://www.astro.uvic.ca/~alan/Nearby_Dwarf_Database_files/NearbyGalaxies.dat}} as  on the 17th of June 2013. 
 Therefore Canis Major
and AXXVII 
 are considered here as dwarf galaxies although  their nature is debated \citep[see][]{Momany04,MartinezDelgado05,Mateu09,Martin16}.
  Most probable satellite position values 
and their corresponding Gaussian width uncertainties in the radial Sun - satellite distance
have been taken from
this database,
 as summarized in 
 \citet{Pawlowski13} Table 2.
The sample analyzed in this paper  considers all the classical plus SDSS satellites.
We will ignore the more recently discovered dwarf galaxy candidates
\citep[see for example][]{DES15,Koposov15a}
which originate from a wide variety of sources/surveys,
 for the sake of consistency with \citet{Pawlowski13}, and for simplicity as far as comparisons are concerned \citep[see also][]{Pawlowski16}. 
  \footnote{ As shown in \citet{Pawlowski14,Pawlowski15b},
    the majority of the recently discovered dwarf galaxy candidates in the MW align with the VPOS. While their consideration in our analysis would produce different results to those presented here, the same conclusions remain regarding the general quality behaviour of the MW planar structures.
    }

The different observational planes of satellites claimed  in the literature 
which we will compare to (defined considering the same sample of satellites)
are listed in \Tab{table_obspl}. This Table shows the  properties (see next section for property definitions)
 of these observed planes as reported by the  most up-to-date studies.  
For the MW these planes are: the so-called classical \citep[i.e., the 11 most luminous MW satellites, see][]{Metz07},  
the VPOSall \citep[][defined by all the 27 confirmed satellites within 300 kpc]{Pawlowski13}, and the
VPOS-3 \citep[][defined by 24 out of 27 of the VPOSall satellites]{Pawlowski13}.
For M31, there is the plane of satellites noted by \citet{Ibata13} and \citet{Conn13}  with the PAndAS survey,
 which we will consider with 14 satellites
 \citep[as analyzed in][hereafter the `Ibata-Conn-14' plane ]{Pawlowski13}\footnote{\citet{Pawlowski13} did not consider AXVI as a satellite because it is further than 300 kpc away.}, and the so-called GPoA \citep{Pawlowski13}, with 19 members (the 14 of the `Ibata-Conn-14' plane plus 5 more).
 We note that other planes of satellites in the Local Group have been suggested in \citet{Shaya13}, but under the consideration of a different initial sample of satellites than that used here. In particular, they define 4 satellite planes (2 in the MW and 2 in M31). The so-called ``plane 1" includes a majority of satellites that   participate in the GPoA, while ``plane 4" is basically a reduced version of the classical plane in the MW plus dwarf galaxy Phoenix.

\section{Searching for predominant planar configurations and plane quality analysis}

Our method to find planar structures and assess their quality consists of 2 parts. The first part follows the 
'4-galaxy-normal density plots' method described in \citet{Pawlowski13}. 
This technique checks if there is a subsample of a given satellite sample that defines 
a dominant
 planar arrangement 
  in terms of the outputs of 
the standard Tensor of Inertia (ToI) plane-fitting technique   \citep[see][]{{Metz07,Pawlowski13}}, 
based on an orthogonal-distance regression.
In terms of the corresponding concentration ellipsoid, planes
are  characterized by:
\begin{itemize}   
\item  $N_{\rm sat}$: the number of satellites in the subsample;
\item $\vec{n}$, the normal to the best fitting plane; 
\item $c/a$: the ellipsoid short-to-long axis ratio;
\item $b/a$:  the ellipsoid intermediate-to-long axis ratio;
\item $\Delta$RMS: the root-mean-square thickness perpendicular to the best-fitting plane;
\item $D_{\rm cg}$: the distance from the center-of-mass of the main galaxy to the plane.
\end{itemize}
These outputs are used to quantify the quality of planes.
To begin with, a planar configuration must be flattened (i.e., low $c/a$), and, as opposed to filamentary, it also requires $b/a \sim 1$  for an oblate distribution.
High quality planes are those
 with a high $N_{\rm sat}$, and a low $c/a$ and $\Delta$RMS,
meaning they  are populated and thin (plane  quality as understood in this work will be specified in more detail at the end of the next section).
 Moreover,  the plane normal $\vec{n}$ determines the plane direction, for example in view of Aitoff projection purposes.  
 Finally,
  a low $D_{\rm cg}$ means that the plane passes near the main galaxy's center; 
  a characteristic to be requested 
if the planes are expected to live within a potential making them  dynamically stable, assuming that
the host galaxy center is close to the center of the system's gravitational potential well.

The second part of our methodology, which is the focus of this paper, is an extension to the
4-galaxy-normal density plot 
method, consisting of a quality analysis of the predominant planar arrangements found.

\subsection{4-galaxy-normal density plot method}\label{N4p}
This method was presented in Sec. 2.4 of \citet{Pawlowski13}. We briefly summarize it and mention the procedure  particularities followed in this study.

\begin{enumerate}

\item 
A plane is fitted to every combination of four\footnote{
Three points always define a plane, not allowing any quantification of plane thickness.
Therefore $4$ is the lowest possible amount to take into consideration under the condition of making the number of combinations as high as possible. This is important in order to analyze sets of satellites consisting of a low number of  objects, as  is frequently the case in satellite populations.
}
 satellites' positions, using the ToI  technique. The resultant normal vector (i.e. 4-galaxy-normal) and corresponding plane parameters are stored.
 To account for distance uncertainties, this step is repeated 100 times using 100 random positions per satellite,  calculated
 using their corresponding radial distance uncertainties.

\item
All the 4-galaxy-normals 
(from all 100 realizations)
 are projected on a regularly-binned sphere, assuming a Galactocentric coordinate system such that the MW's disc spin vector points towards the south pole. 
A density map (i.e. 2D-histogram)  is drawn from the projections, where 
each normal has been weighted by $\log\left( \frac{a+b}{c} \right)$ to emphasize planar-like spatial distributions.
The over-density regions   in these density maps  (i.e. regions of  4-galaxy-normal accumulation) 
therefore signal the normal direction to 
a dominant planar space-configuration. Satellites contributing 4-galaxy-normals to a given over-density are likely members of such a dominant plane. 
As opposite normal vectors indicate the same plane, density maps  in this study are shown 
through Aitoff  spherical projection diagrams in Galactic  coordinates (longitude $l$, latitude $b$) within the $l=[-90^\circ,+90^\circ]$ interval.

\item
We order bins   by density value. The main over-density region is identified 
around
the highest value bin. Subsequent  over-densities are identified by selecting the next bin,
in order of decreasing density,
 which is separated more than 15$^\circ$  from the center of all the previously defined over-densities. In this way
over-density regions  
  are differentiated and isolated. 
  For each of these regions, the midpoint of the   highest-density bin  will define the corresponding 
  \textit{ density peak}'s  coordinates.

\item
We quantify how much a certain satellite $s$ has contributed to a given density peak $p$ (which we refer to as  '$C_{p s}$'). 
To this end, we define an aperture angle of 15$^\circ$  around the density peak, selecting all 4-galaxy-normals within it. 
For each of them, the four contributing satellites  are determined.
A given satellite $s$ is counted to contribute  the 4-galaxy-normal's \textit{weight} to peak $p$.
Therefore, its final contribution $C_{p s}$, is 
the sum of weights corresponding to  all the  4-galaxy-normals within the peak aperture that satellite $s$ contributes to.
Finally,  all satellites are  ordered by decreasing  $C_{p s}$ to the density peak $p$, such that the first satellite is that which contributes most.

We note that changing the bin size used in our analysis does not modify our results, as we find the same overdensity regions.
While it does slightly change the position found for the density peak centres, the differences are small and do not modify the final order of satellites by  $C_{p s}$. Therefore the final results remain unaltered.

\end{enumerate}

\subsection{An extension to the method: plane quality analysis}
\label{PMExt} 

To allow an individual and in-depth analysis of each overdensity, we present an extension of the  4-galaxy-normal density plot method.
Rather than a plane per overdensity, the extension will provide us with a collection of planes with a different number of satellites.

For each over-density $p$ 
we initially fit a plane to the $N_{\rm sat}$=7  satellites with highest $C_{p s}$ 
(i.e., the 7 satellites that contribute most  to 4-galaxy-normals within  15$^\circ$ of the density peak),
 and store the resultant plane parameters. 
This number $N_{\rm sat}$=7  is low enough to allow for an analysis of ToI parameter behaviour as $N_{\rm sat}$ increases, and at the same time high enough that we begin with populated  planes. Note that taking instead $N_{\rm sat}=7 \pm 2$  to begin with does not alter our conclusions.

Then, the next satellite in order of decreasing 
$C_{p s}$
 is added to the group of satellites. Again a plane is fitted to their positions and the parameters stored.
This plane-fitting process is repeated
until all contributing satellites are used.

To include the effect of distance uncertainties, in practice we calculate 1000 random positions per satellite, and fit 1000 planes at each iteration with $N_{\rm sat}$ satellites.
The final results at each $N_{\rm sat}$ 
 correspond to the mean values from these random realizations  and the corresponding  errors to the standard deviations.

In this way, for each over-density found  we obtain a  \textit{collection} or catalog of
planes of satellites, each plane consisting of an increasing number of members,
as well as the quality indicators for each of them.

In this work ``high quality" means  populated and
flattened planes.
This is quantified through $N_{\rm sat}$ and 
$c/a$ 
\citep[and/or $\Delta$RMS,
note that they are very often correlated; see][]{Pawlowski14b}. 
 Being a two-parameter notion, to  compare 
planes' qualities  we need that  either $N_{\rm sat}$ is constant or that $c/a$ is constant (or that at least it varies very slowly with $N_{\rm sat}$). 
In the first case, lower $c/a$ means higher quality. In the second case, more populated planes are rated as of higher quality.  Another case when comparison 
is possible is when one plane is more flattened and populated than another: the first has a higher quality than the second. 
These considerations have been applied to the different member planes in the collection obtained for each density peak, allowing us to make quality comparisons, 
in particular with already determined planes,  and, very interestingly,  to single out new high-quality ones.
\footnote{  Note that quality as understood in this paper should not be confused with `significance' (meaning the frequency of planes with given characteristics in randomized satellite systems). }

\section{Results}

\begin{figure*}
\centering
\includegraphics[width=0.49\linewidth]{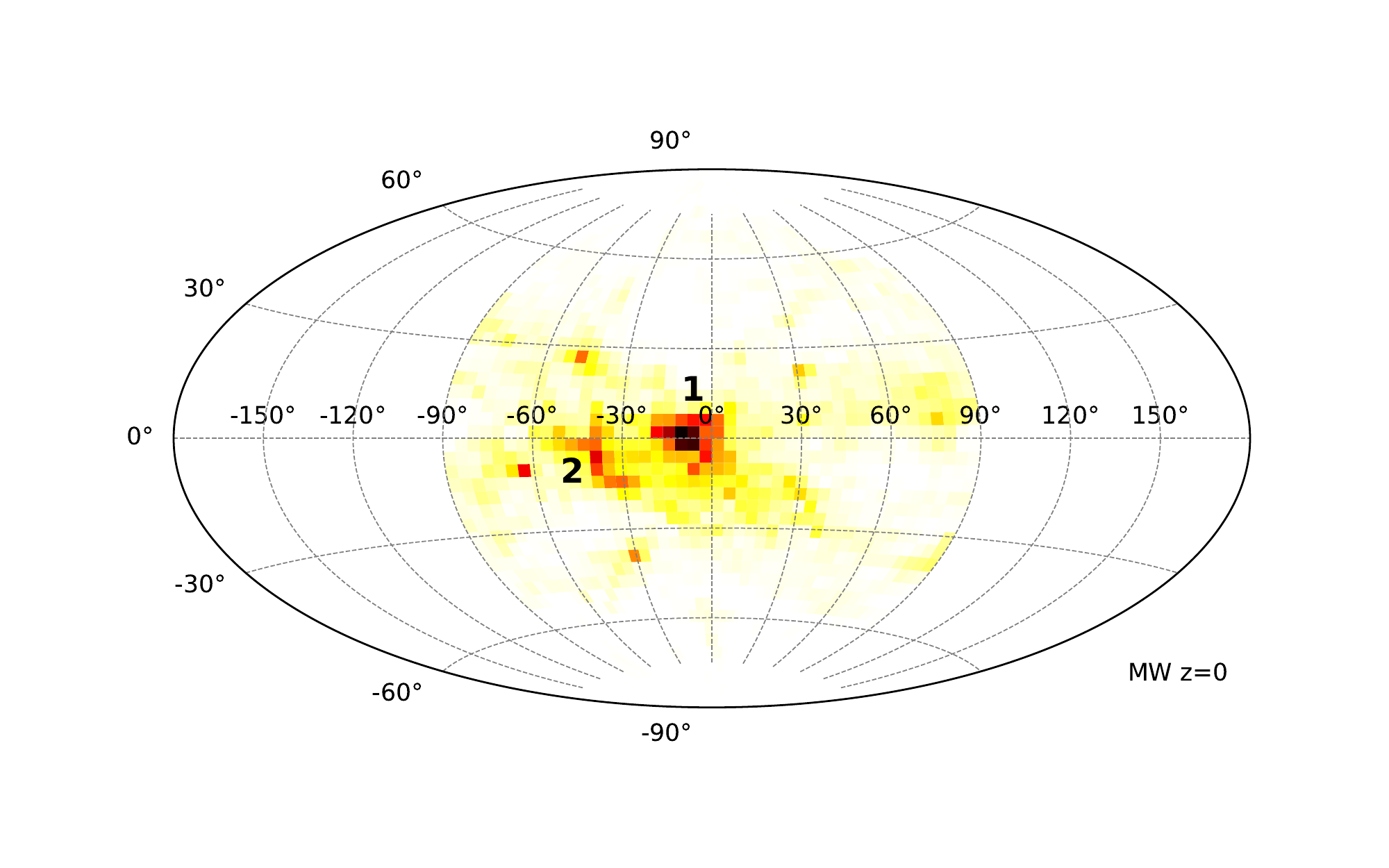}
\includegraphics[width=0.49\linewidth]{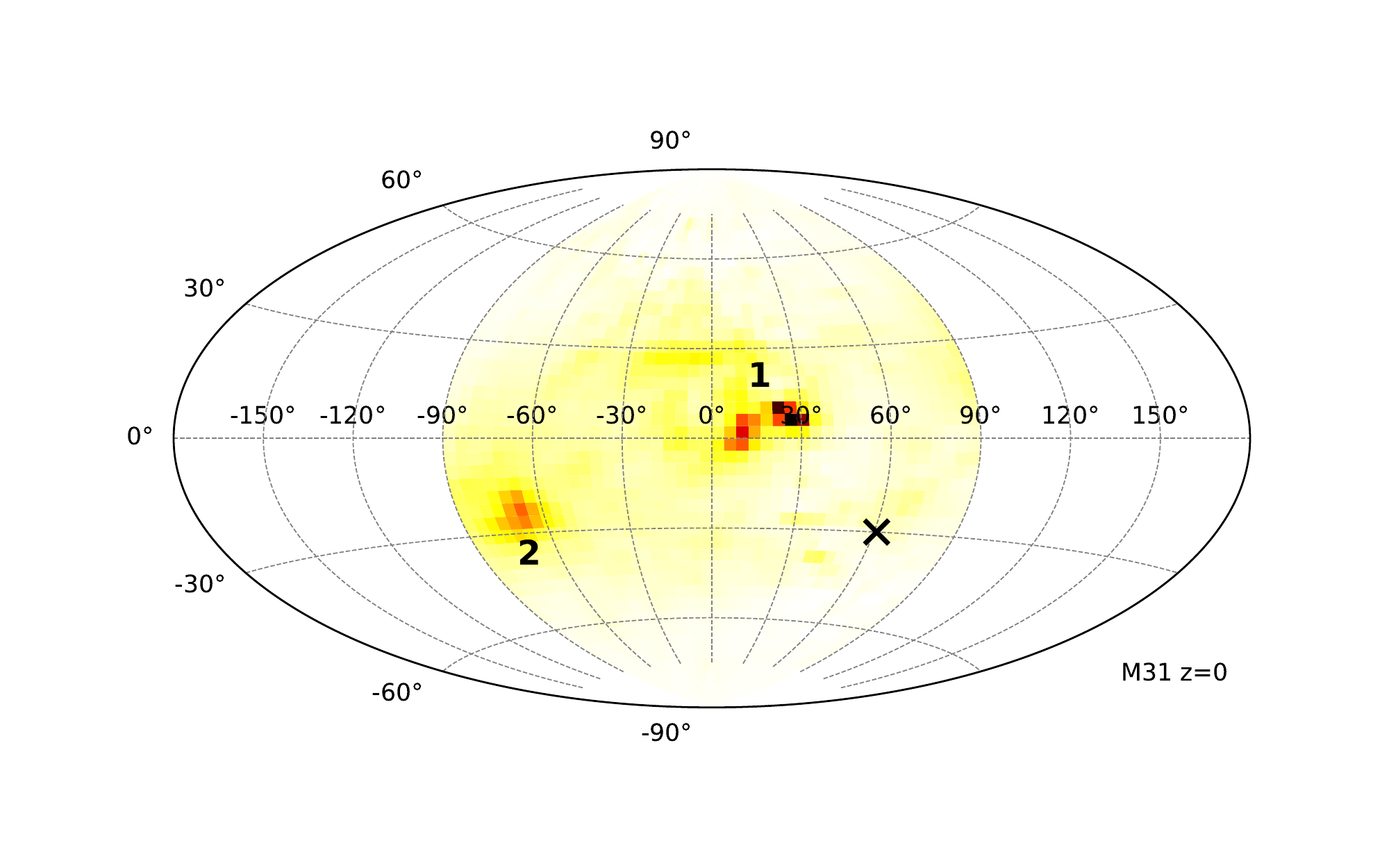}
\caption{Aitoff projection diagrams of the Milky Way (left) and M31 (right) 4-galaxy-normal density plots
\citep[see also Figs. 2 and 4 in][]{Pawlowski13}.
The colormap shows the 
number of 4-galaxy-normals within a bin, each weighted by $\log\left( \frac{a+b}{c} \right)$
to emphasize planar-like spatial configurations
(see \Sec{N4p}
for details).  
The total number of 4-galaxy-normals  is 1755000   for the MW  and  4637600 for M31, taking into account 100 random realizations for each (see text).
The relevant over-density regions in each map are labeled in order of intensity (Peaks 1 and 2 in the main text).
Each diagram is centered on its corresponding central galaxy but the   orientation of coordinates in both cases is such that the disc of the MW lies on the latitude $b=0^\circ$ plane and its spin points along the OZ axis.
M31's spin is marked with an \textbf{X}.
A much finer bin size than that shown here has been used to extract the density peak coordinates.
}
\label{MWM31dp}
\end{figure*}

\Fig{MWM31dp} shows  Aitoff projection diagrams of the  4-galaxy-normal density plots obtained for the MW and M31 satellite systems. 
 These Aitoff diagrams can be compared to  the contour plots in   Figs. 2 and 3 from  \citet{Pawlowski13}, to which they are essentially identical\footnote{
{The conversion from the  galactocentric longitude convention used here ($l$) and that used  in \citet{Pawlowski13} ( $l'$) is: $l = l' -180 (^\circ)$}.
}.

\begin{figure*}
\centering
\includegraphics[width=\linewidth]{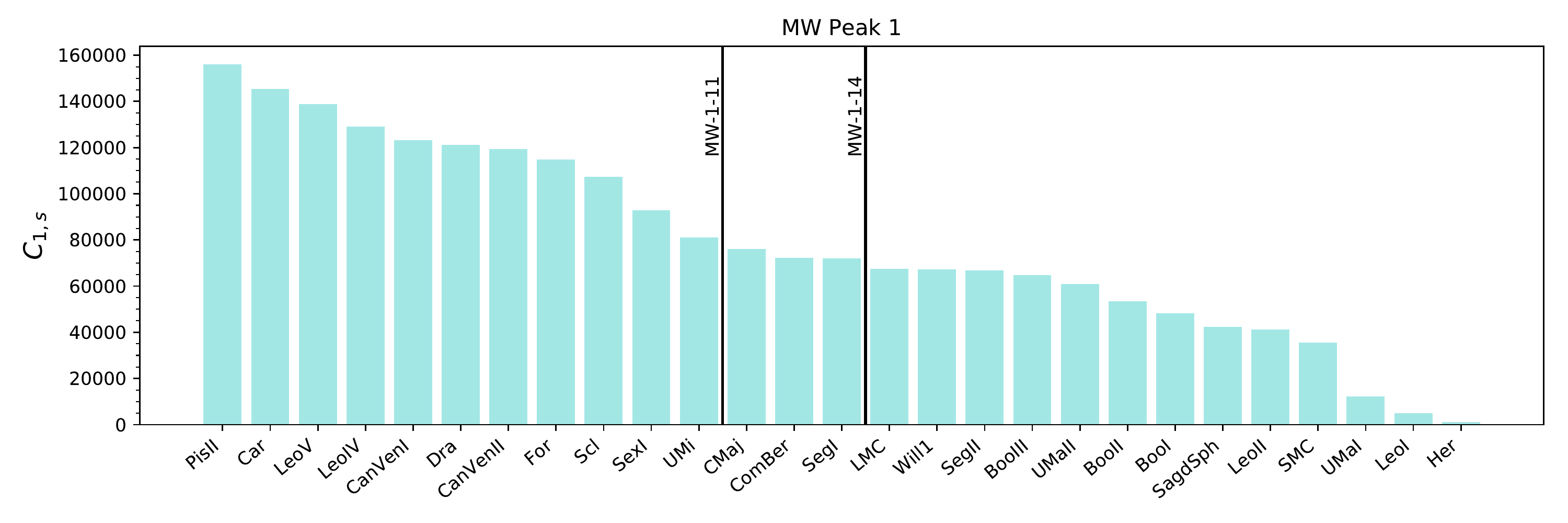}
\includegraphics[width=\linewidth]{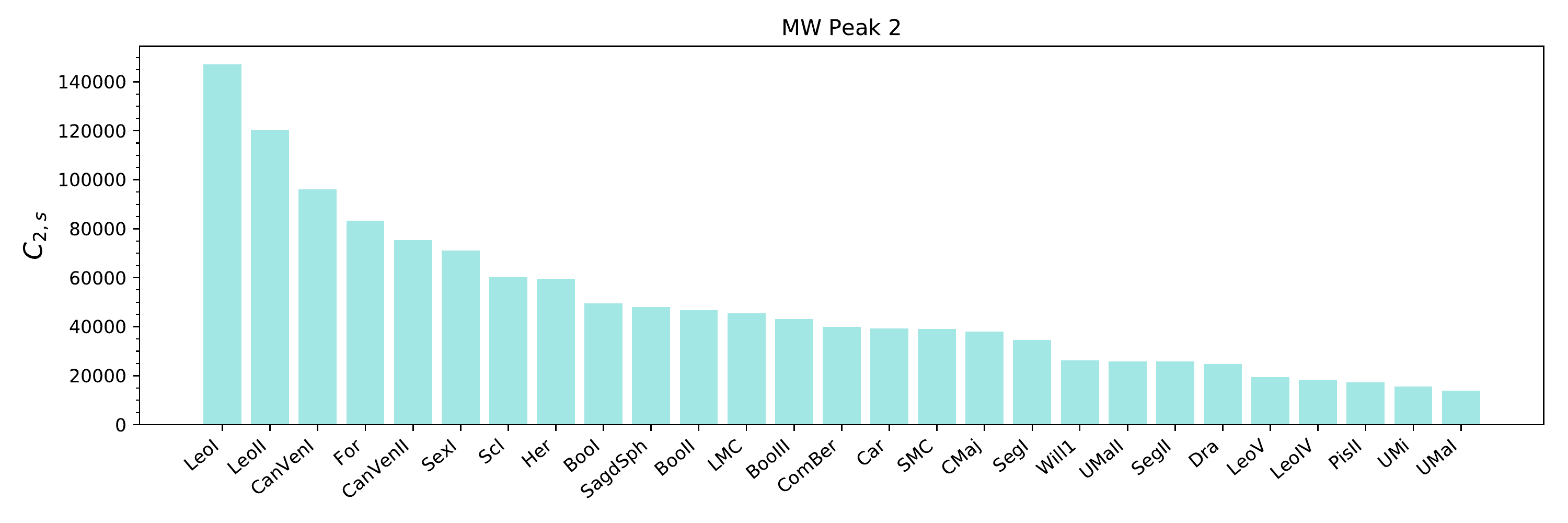}
\caption{Bar chart showing the contribution 
of  satellites to 4-galaxy-normals in 15$^\circ$  around the first  ($C_{1s}$, top panel) and second ($C_{2s}$, bottom panel) most important  over-densities found in the Milky Way density plot (left panel \Fig{MWM31dp}).
The sets of objects that make up the planes of satellites singled out in this work are delimited with vertical lines and labeled correspondingly (see Tab \ref{table_obsThisWork}).
}
\label{barMW}
\end{figure*}

\begin{figure*}
\centering
\includegraphics[width=\linewidth]{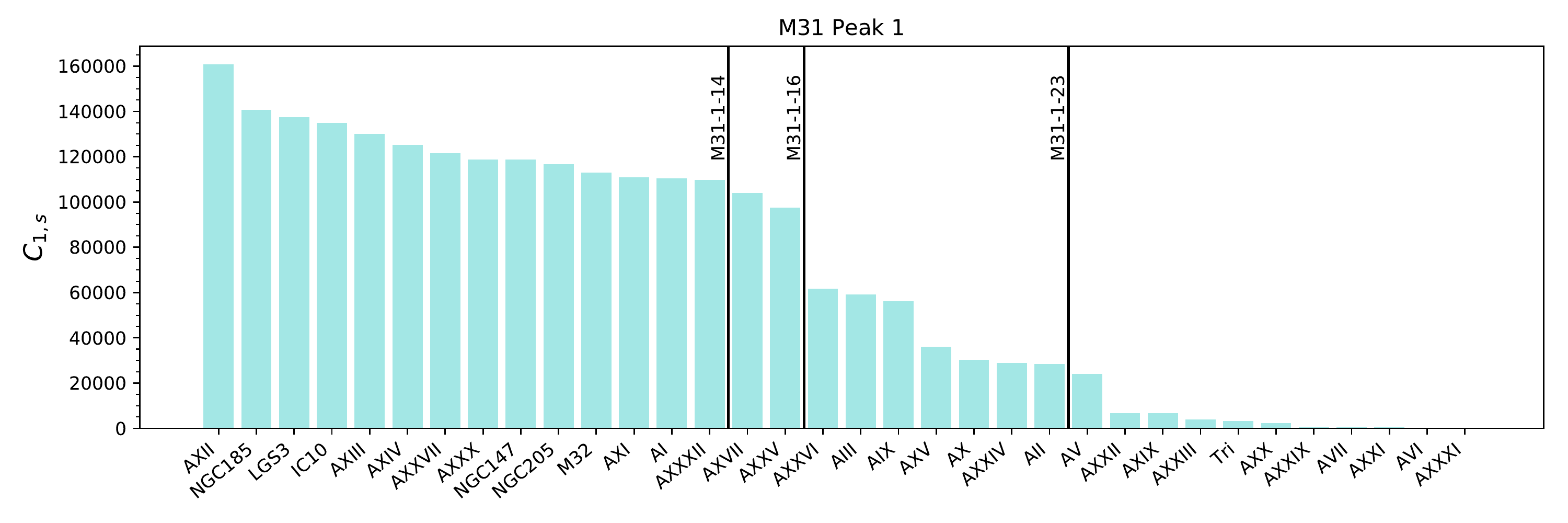}
\includegraphics[width=\linewidth]{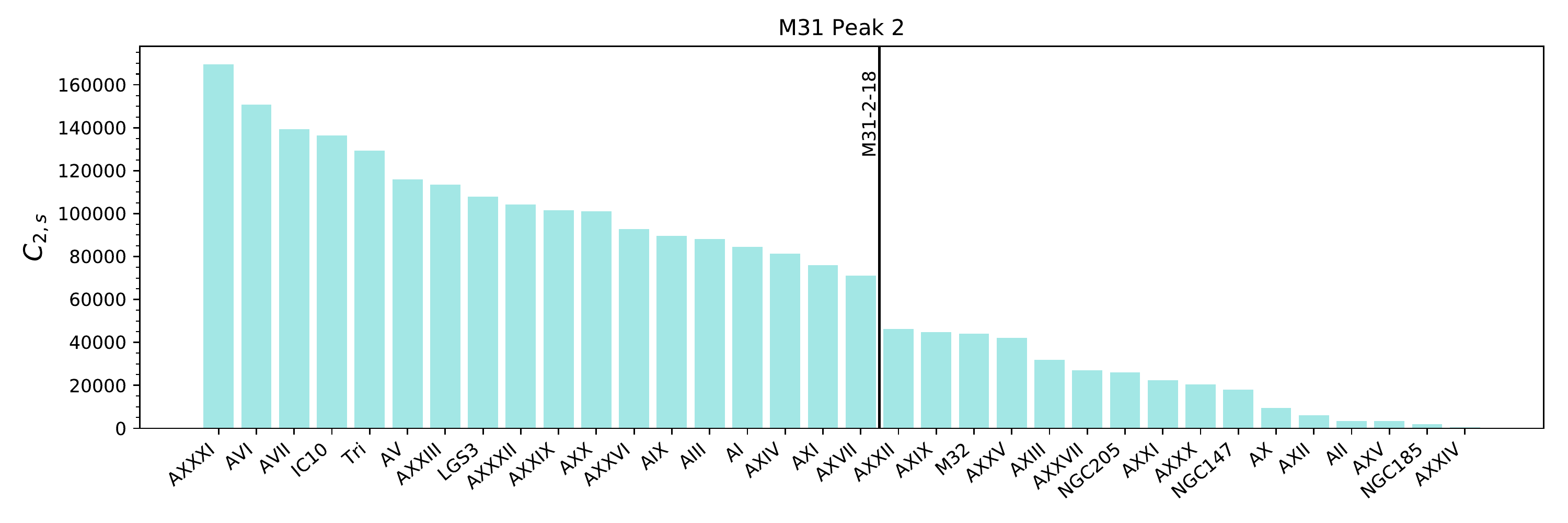}
\caption{Same as \Fig{barMW} for M31. Both the first (top panel) and second (bottom panel) over-densities of M31's 4-galaxy-normal density plot have similar intensities which is reflected in a high and comparable contribution from satellites in both bar charts.
}
\label{barM31}
\end{figure*}

\subsection{Milky-Way}\label{MW}
As reported in \citet{Pawlowski13},
the MW density plot shows that 4-galaxy-normals are mainly clustered in the region central to the diagram,
revealing a planar structure that is polar to the Galaxy (i.e., the normal vector to the plane is perpendicular to the Galaxy's spin vector).
There is one dominant over-density (Peak 1)
located at 
 $(l,b)=(-10.23, 0.41)$,
 and  a second lower density peak (Peak 2) close to the first
 at
 $(l,b)=(-39.68, -4.50)$. We will neglect the few isolated bins with intermediate intensity.

Figure \ref{barMW} shows satellites ordered by contribution
to 4-galaxy-normals within 15$^\circ$ around both density peaks  (i.e., $C_{1s}$, upper panel, and  $C_{2s}$, bottom panel).  
There are 10 satellites  that dominate the contribution
 (i.e., $C_{p s} > 0.5 \times {\rm max}(C_{p s}, s = 1, ..., 27$)), 
 in Peak 1 (i.e., PisII, Car, LeoV, CanVenI, LeoIV, Dra, For, CanVenII, Scl, SexI),
  while 6    contribute most in Peak 2 (i.e., LeoI, LeoII, CanVenI, CanVenII, For, SexI)\footnote{Both the location of these peaks  and their major contributers are consistent with the results reported in  \citet{Pawlowski13} (see their Fig. 3).}.
  In this case, the contribution is mainly driven by LeoI and LeoII, while the rest of satellites are common with Peak 1 and take low $C_{p s}$ values, indicating the low relevance of this second structure.

\subsubsection{Quality analysis}

Following our extension to the 4-galaxy-normal method  (\Sec{PMExt}),  for each over-density region
  we have iteratively computed planes of satellites with an increasing number of members $N_{\rm sat}$,
  following the order of satellites given in Fig. \ref{barMW}.
In this way,
 for each peak we obtain an ordered collection of planes, one plane for each $N_{\rm sat}$. 
These collections, as well as the identities of the satellites belonging to each plane, can be read out of Figure \ref{barMW}, for both peaks.

\begin{figure*}
\centering
\includegraphics[width=0.49\linewidth]{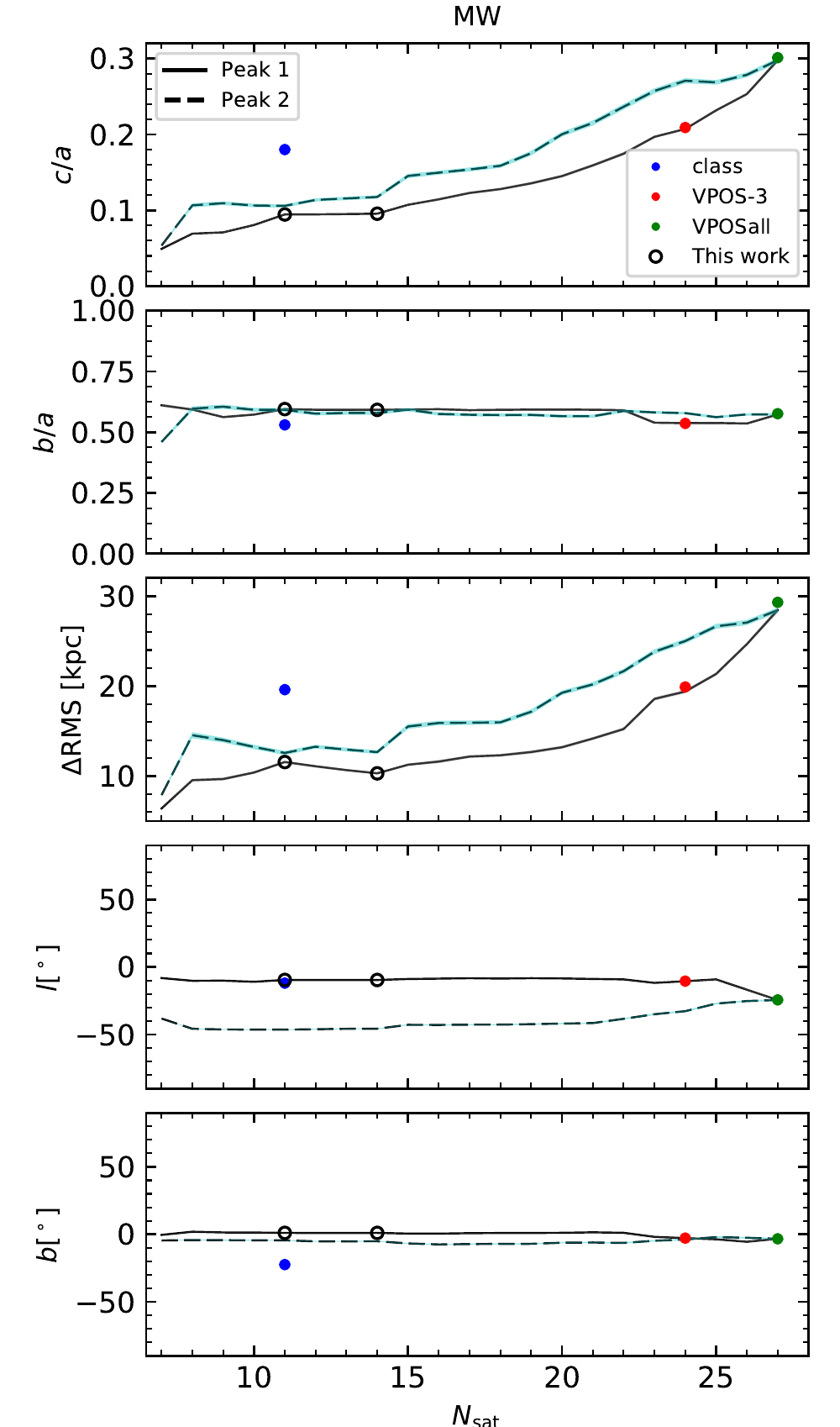}
\includegraphics[width=0.49\linewidth]{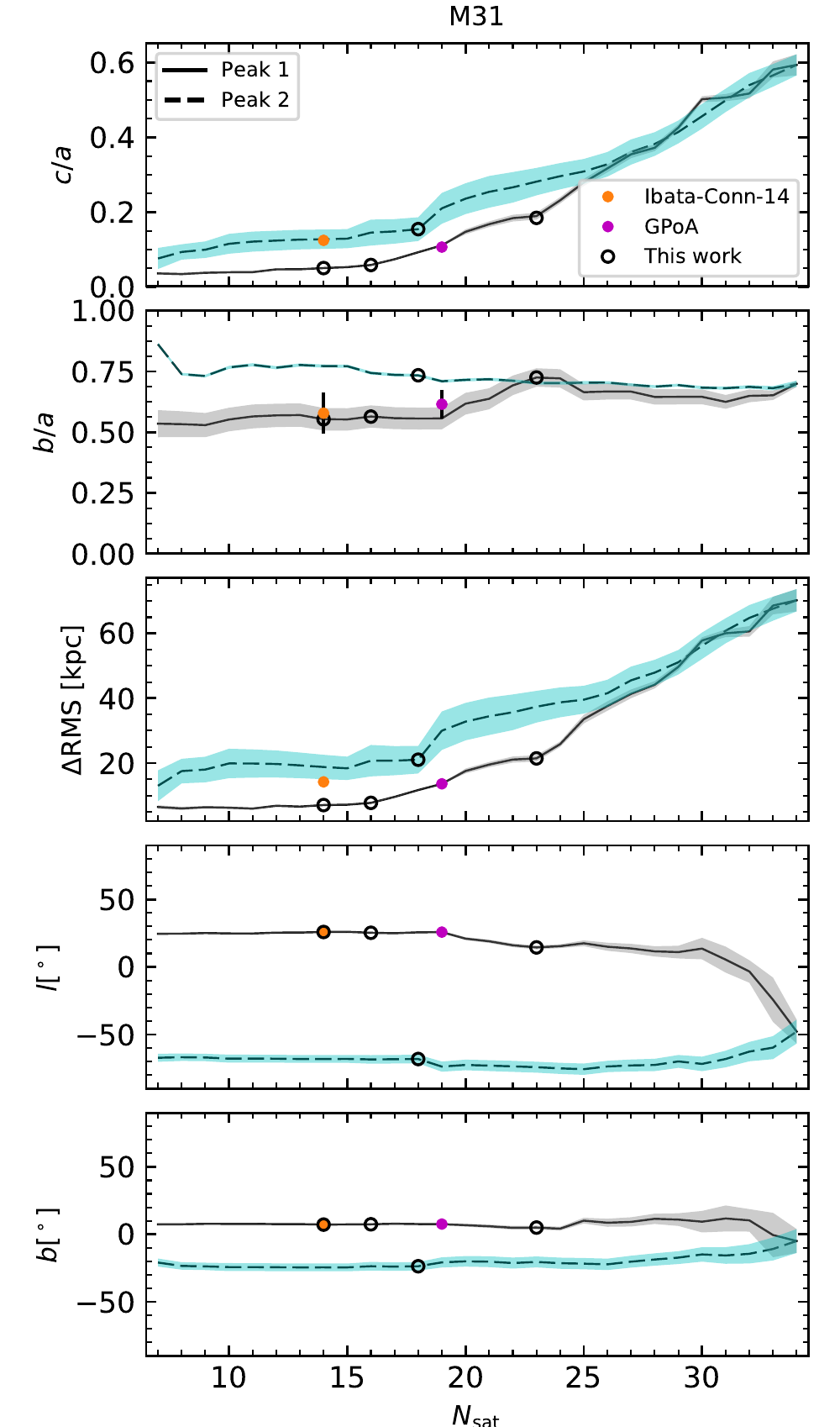}
\caption{Quality analysis of the main planar structures found in the Milky Way (left panels)
and in M31 (right panels)
with the 4-galaxy-normal density plot method (see Figs. \ref{MWM31dp}).
Lines show
  $c/a$, $b/a$, $\Delta$RMS and 
the direction of normal vectors to the best-fitting plane $(l,b)$ as a function of $N_{\rm sat}$.
 Except in the  $l$ or $b$ versus $N_{\rm sat}$  panels, 
results are the mean values of 1000 realizations at each $N_{\rm sat}$, and shaded regions show the standard deviations.
 $(l,b)$   directions have been calculated using the most-likely positions of satellites  and shaded regions show the corresponding uncertainties
in terms of the spherical standard distance $\Delta_{\rm sph}$ \citep{Metz07}.
As expected, the two curves in each panel converge for the maximum $N_{\rm sat}$ since the samples of satellites become identical.
Colored circles
show the results for the reported observed planes of satellites in the MW (classical, VPOS-3, VPOSall),
 and in M31 (`Ibata-Conn-13' and GPoA), respectively.
 Their specific values including their errors are given in \Tab{table_obspl}.
The planes of satellites singled out in this work are shown with black open circles, and their corresponding  ToI parameters 
are given in \Tab{table_obsThisWork}.
}
\label{MWM31_ca_frac}
\end{figure*}

The left panel of 
\Fig{MWM31_ca_frac} shows the results obtained for Peak 1 (solid line) and Peak 2 (dashed line) in the MW.
The collection of planes obtained for a given over-density region gives rise to a set of points, 
with their corresponding error bars, one at each $N_{\rm sat}$ value.
 They are shown joined with a line, with the corresponding error bands.
 These are very narrow in the case of the MW, showing that the MW results are not that affected by distance uncertainties.

First we see that, for any $N_{\rm sat}$ value, $b/a$ is rather high (and constant), while $c/a$ is low. Therefore configurations are indeed  \textit{planar-like}. Moreover,
the  lines for both density peaks show rather smooth trends of increasing $c/a$ and $\Delta$RMS with increasing $N_{\rm sat}$.
In particular,
 the MW Peak 1 line 
 defines a planar structure  of  satellites with a higher quality (i.e., lower $c/a$,  lower $\Delta$RMS  at given $N_{\rm sat}$), than that defined by Peak 2. 
This is expected, given the higher density of 4-galaxy-normals in Peak 1 than in Peak 2
(see Fig. \ref{MWM31dp}).
This suggests
that the MW satellite sample seems to be a unique highly planar-like  organized  structure, as we had already learnt from Fig. \ref{barMW}.
We also note that both the $c/a$  and $\Delta$RMS   versus $N_{\rm sat}$  curves  roughly show  the same shape patterns, 
due to the lack of significant $b/a$ variation as   $N_{\rm sat}$ increases. Therefore, including $\Delta$RMS on top of $c/a$ in the quality analysis
generally  does not add any relevant information.

In the bottom panels we give the directions of the normal vectors corresponding to the best fitting planes (obtained from the satellites' most-likely positions) as a function of $N_{\rm sat}$. Shaded regions show the
 corresponding spherical standard distances $\Delta_{\rm sph}$ \citep{Metz07},
a measure of the planes normals' collimation 
  for the 1000 realizations.
The normal directions to planes from both Peak 1 and 2 in the MW remain very stable as  $N_{\rm sat}$ increases, and their uncertainties due to satellite distance errors are very small.
   
As for the plane distances to the MW, $D_{\rm cg}$ is  below 16 kpc in all plane-fitting iterations in Peak 1, and below 12.5 kpc in the case of Peak 2.

For reference, overplotted
colored points show the values on this diagram for
the  observed planes of MW satellites mentioned in the literature (classical, VPOS-3, VPOSall; see \Tab{table_obspl}),
defined from Peak 1 \citep{Pawlowski13}.
We note that the points corresponding to the VPOS-3 and VPOSall planes  fall over the trend given by the solid  line, as expected.
  On the other hand, and very interestingly, 
 this analysis    shows that 
    there  is  a  different 
 combination of $N_{\rm sat}$=11 satellites that results in a much flatter and thinner plane than the classical one
(see MW-1-11\footnote{Planes underlined in this work, either for the MW or M31, are named after the peak where they have been identified and the number of satellites they include (i.e., [Host-Peak-$N_{\rm sat}$]). } in Table \ref{table_obsThisWork}).
 In fact, the plane including $N_{\rm sat}$=14 satellites (MW-1-14 in Table \ref{table_obsThisWork})
presents an even higher quality than that with $N_{\rm sat}$=11, as  
$c/a$  remains  roughly constant at a higher $N_{\rm sat}$.

\begin{figure}
\centering
\includegraphics[width=\linewidth]{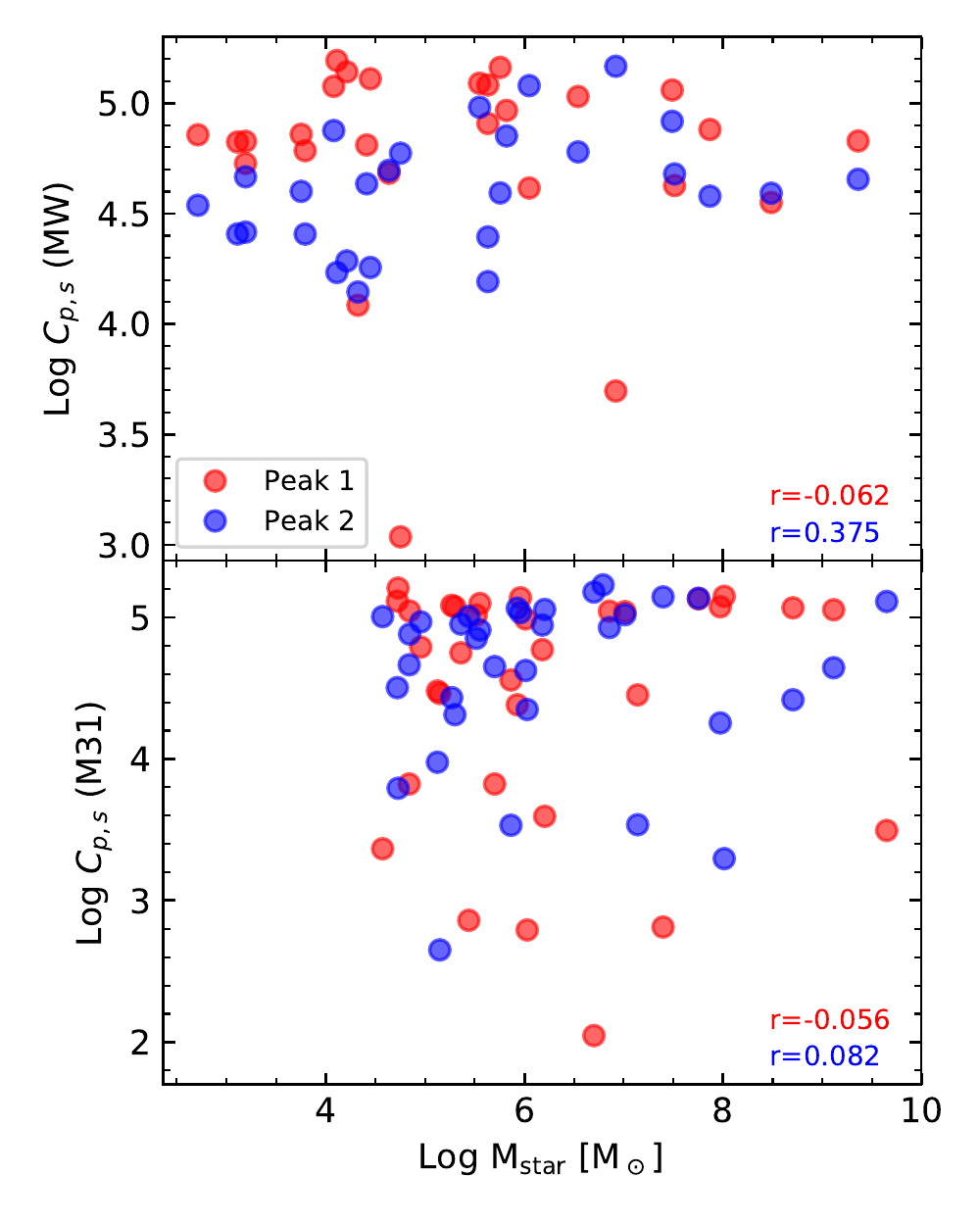}
\caption{
The contribution of satellites to 4-galaxy-normals within 15$^\circ$ of the main density peaks, $C_{p,s}$, versus stellar mass. The Pearson correlation coefficients $r$ 
are given in each case. 
}
\label{cps_mstar}
\end{figure}

This is possible
because this analysis uses  the 3-dimensional information of positions, while the classical plane of satellites
 was found observationally
  when only the most luminous (i.e. massive) satellites were known to exist.
   This important result
 indicates
 that planes of satellites are not necessarily  composed by the most  massive satellites of a galactic system
  \citep[see also][]{Libeskind05,Collins15} 
  and hence they should not be searched for in this way.
  Indeed, we find that $M_{\rm star}$\footnote{
Observational stellar masses have been computed applying the \citet{Woo08} mass-to-light ratios to the luminosities in \citet{McConnachie12}.} 
is not correlated with $C_{1s}$ (contribution to the main over-density region, where we find the highest quality planes):
 the correlation coefficient $r$  is low, in such a way that the probability of getting such value assuming that there is no correlation is higher than 76\% (see upper panel in Fig. \ref{cps_mstar}).

\subsection{Andromeda M31}\label{M31}

Figure \ref{MWM31dp} reveals,  as in the MW,
 two 4-galaxy-normal  over-densities in M31.
In this case they both show high, comparable intensities and are located quite separate ($\sim80^{\circ}$) from one another.
We define Peak 1 
as the over-density at
  $(l,b)=(26.60, 6.14)$, 
 and Peak 2 as that located at  
 $(l,b)=(-69.14 , -21.68)$. 
The direction of
 M31's spin vector, depicted with an 'X', indicates that
the planar configurations defined by both peaks
  are not perpendicular to the galaxy's disc, but inclined 
$48.93^\circ$ and $70.49^\circ$, respectively.
Interestingly,
Peak 1 forms an  angle of $83.86^\circ$ with the MW's spin vector, meaning its corresponding planes are approximately normal to the MW's disc
 \citep{Conn13}. 
Furthermore, the projected angular distance on the sphere between Peak 1 and Peak 2 is of
 $82.43^{\circ}$,
 and the angle between Peak 1 (Peak 2)  and the Sun - M31 line is $87.73^{\circ}$ ($5.31^{\circ}$). Therefore the
planar configuration of satellites defined by Peak 1  
  is observed nearly edge-on from the MW (see also \citealt{Pawlowski13} Table 5), 
while that of Peak 2   is approximately perpendicular to it and would be observed mostly face-on.

Fig. \ref{barM31} shows that approximately half of the satellite sample contributes dominantly to each corresponding peak $p$.
Focusing on the identities of satellites, we find that,
out of the 16 satellites with highest $C_{1 s}$, 9 are among the satellites with lowest $C_{2 s}$.
On the other hand,
out of the 17 satellites with highest $C_{2 s}$, 10 are among those with lowest $C_{1 s}$.
This is indicating that the two over-densities' contributing members are not the same.
While  satellites contributing most to Peak 1 define the GPoA plane from \citet{Pawlowski13}
(and also generally coincide with satellites in ``plane 1" from \citet{Shaya13}),
Peak 2 and its corresponding predominant planar satellite configuration have not been analyzed prior to this study\footnote{Note that the  planar structure derived here from Peak 2 does not correspond to the so-called 'M31 disc plane' noted in \citet{Pawlowski13}, or to ``plane 2"  in \citet{Shaya13}. }
and will be described in detail below.

\subsubsection{Quality analysis}

For each peak, 
we build up their respective collections of planes of satellites by
 iteratively applying the best-fitting 
 plane technique to an increasing number of satellites  $N_{\rm sat}$
following the order given by Fig.  \ref{barM31}.
The values of concentration ellipsoid parameters versus $N_{\rm sat}$ are shown in  the right panel of Fig. \ref{MWM31_ca_frac}.
We see that  $c/a$ and  $b/a$ take respectively low and high values, confirming that these spatial distributions are actually planes up to $N_{\rm sat} \sim 24 - 25$. When considering the whole sample of M31 satellites, $c/a$ and $b/a$ take a similar value of $\sim0.6$, indicating instead a non-flattened ellipsoidal spatial distribution.

The corresponding parameter error bands are clearly apparent in the case of M31 as compared to the MW, due to overall larger satellite distance uncertainties.
As mentioned previously, 
while the GPoA is viewed approximately edge-on from the Sun, the planes of satellites from Peak 2 
 are viewed mostly face-on. Therefore the uncertainties in
the Sun - satellite distances affect more (less) to the $c/a$ and
$\Delta$RMS parameters 
than to $b/a$, in the case of Peak 2 (Peak 1). 
This fact explains the different magnitude of the error bands among the ToI parameters shown in this figure.

Focusing on the $c/a$ and $\Delta$RMS panels, one can see that
only up to $\sim$half 
of the total number of satellites contributing  respectively to Peak 1 and Peak 2
form a thin planar structure, which rapidly thickens  as more members are added to the plane-fitting iteration.
Moreover, as said above, satellite identities contributing most to both peaks are overall different as shown in \Fig{barM31}.
Therefore,  in contrast to the MW, the M31 satellite  sample does not form one preferential planar structure but seems to be divided in (at least) two.

The normal directions to the planes are  stable as $N_{\rm sat}$ increases and reaches  $N_{\rm sat}=$19
for Peak 1 and $N_{\rm sat}$=18 for Peak 2, showing that the two satellite planar structures are well defined. 
Beyond these values, the normals to the corresponding planes are not that well fixed.

As for the distances, planes around Peak 1 pass close to the M31 center, while planes belonging to Peak 2 do not 
(see values in Tab. \ref{table_obsThisWork}).
Their distances are large
but still within reasonable ranges 
to allow for  the possibility of dynamical stability
in a complex, binary system like the Local Group,
especially if we take into account the large distance uncertainties (see errors in Tab. \ref{table_obsThisWork})
and consider the important lopsidedness in the distribution of M31 satellites \citep[see][]{McConnachie06,Ibata13}.

The specific   values for M31 Ibata-Conn-14 and GPoA planes
are shown with
colored circles in Fig. \ref{MWM31_ca_frac} (see \Tab{table_obspl}).
 Our methodology reveals a combination of $N_{\rm sat}$= 14 satellites (marked with a black open circle in Fig.
\ref{MWM31_ca_frac}, right panels) that yields a higher quality plane than the `Ibata-Conn-14' plane
(see M31-1-14 entry in  \Tab{table_obsThisWork} and Fig. \ref{barM31} for satellite identities).  
  This occurs because the  `Ibata-Conn-14' plane was defined among only PAndAS survey satellites,
while the sample used 
in \citet{Pawlowski13} and 
here includes
the PAndAS satellites 
within 300 kpc of M31 (25 out  of 27)
plus 
9  satellites discovered differently
(i.e., LGS3, IC10, AXXXII, AVII, AXXIX, AXXXI, AVI, NGC205, M32).
Interestingly, the latter turn out to be   precisely among the dominant 4-galaxy-normal contributers to both M31 Peaks 1 and 2.

In turn, the magenta circles  corresponding to the GPoA match the solid line (Peak 1)
because the 19 satellites that we find with highest $C_{1s}$ are precisely  the GPoA satellite sample.  
Note that the satellite system shows low correlation coefficients $r$  between 
$M_{\rm star}$ and $C_{1s}$ or $C_{2s}$,
with a  higher than 75\%  probability of getting such $r$ values assuming that there is no correlation (see lower panel of Fig. \ref{cps_mstar}).

Moreover, our analysis allows the identification of the highest quality planes in M31 as  $N_{\rm sat}$ increases at $c/a$ roughly constant.
These planes correspond to the points in Fig. \ref{MWM31_ca_frac} at which the M31 Peak 1 and Peak 2 lines start to
increase rapidly  in the $c/a$ and $\Delta$RMS panels,
 marked in Fig. \ref{MWM31_ca_frac} with
black open circles.
 For Peak 1 the highest quality plane at low $c/a$ occurs with $N_{\rm sat}$=16 (M31-1-16  in Table \ref{table_obsThisWork})
and
for Peak 2,
with $N_{\rm sat}$=18 members (M31-2-18), 
beyond which the normal vector $\vec{n}(l,b)$ to the best-fitting plane does not conserve its direction.
The latter presents  comparable properties to the GPoA (magenta circle),
 to which it is roughly perpendicular. 
Indeed, given  that the GPoA has one satellite more and a lower $c/a$ value than the M31-2-18 plane (see Tabs. \ref{table_obspl} and \ref{table_obsThisWork}), strictly speaking the former
 has a higher quality than the latter. However, the differences are a 5\% in $N_{\rm sat}$, and a 10\% in the $c/a$ values,
 taking into account the error bars. 
Therefore we can conclude that the qualities of both planes are comparable.

\begin{figure}
\centering
\caption{ Edge-on view of  M31-2-18,
the highest-quality plane from Peak 2 in M31  with $N_{\rm sat}$=18 members (in red),
and  the GPoA (in blue), showing their relative orientation.
M31's galactic disc and spin vector are depicted in green. 
Satellites belonging only to the M31-2-18 plane are shown as red points, while satellites belonging only to the GPoA are shown as blue points. Satellites shared by both samples are violet.
}
\includegraphics[scale=0.5]{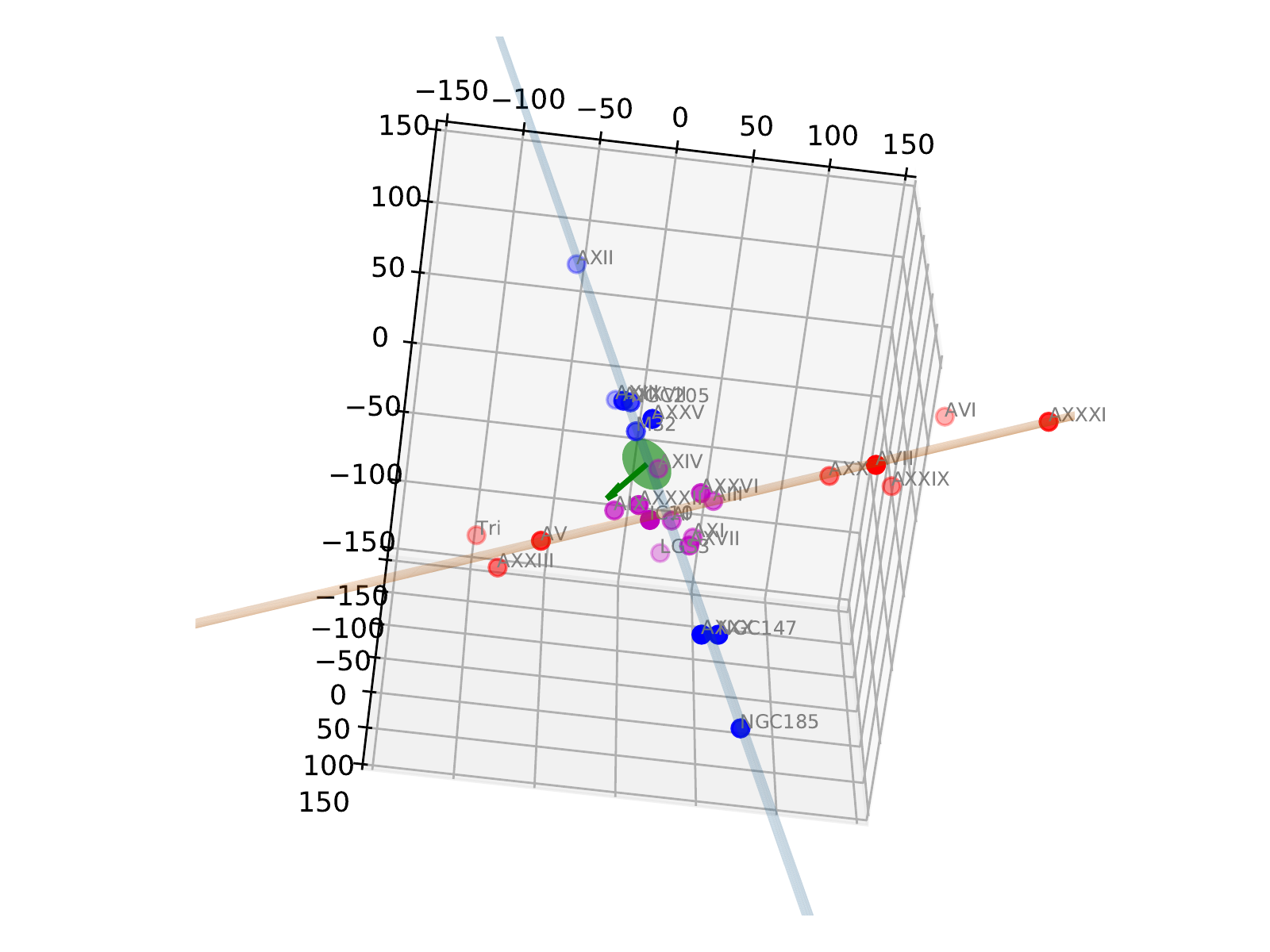}
\label{3D_M31_p1p2}
\end{figure}

Figure   \ref{3D_M31_p1p2} shows the relative orientation 
between M31-2-18 and the GPoA.
Note that 10  satellites are shared by both samples (i.e., LGS3, IC10, AXIV, AXI, AXXXII, AI, AXVII, AIX, AIII and AXXVI, in violet in the figure).

It is of interest whether this plane could be dynamically stable, this is, if the member satellites corotate within the planar structure they define in space. Line-of-sight velocities of the M31-2-18 satellites \citep[taken from][]{McConnachie12,Collins13,Martin14}, which we observe face-on from the MW, give a perpendicular velocity dispersion of $\sigma=90.20$ km/s. 
 According to \citet{Fernando17}, such a plane will be erased in a short timescale and is just a fortuitous alignment of satellites, as they find that 
planes with a perpendicular velocity dispersion above $\sim50$ km/s disperse to contain half their initial number of satellites in 2 Gyrs time.

Finally, we also note the high quality of the plane with $N_{\rm sat}$=23 from Peak 1 (M31-1-23 in \Tab{table_obsThisWork}), with 
 ToI parameter values  very similar to
  those of the VPOS-3 plane of satellites  in the MW.

\section{Summary and conclusions}

Recent studies on planes of satellites have resorted to evermore refined methods to define and characterize them. 
In this work, we have further developed one such method,  the '4-galaxy-normal density plots' method \citep{Pawlowski13}, 
with an extension designed to 
identify, systematically catalog and 
  study in detail the \textit{quality} of the predominant planar configurations 
 revealed by  over-densities in the  4-galaxy-normal density plots.

We count the weighted number of times each satellite $s$ contributes to 4-galaxy-normals within 15$^\circ$ of a  specific over-density $p$ ($C_{ps}$).
For each relevant over-density, we order satellites   by decreasing $C_{ps}$ and 
iteratively fit planes to subsamples of satellites following this order.
In this way, rather than a plane per over-density,
we yield a catalog or \textit{collection of planes} of satellites  with an increasing number of members $N_{\rm sat}$,
 whose normals cluster around the density peak.

 The quality of planes is quantified through the number of member satellites $N_{\rm sat}$ and the degree of flattening. The latter is measured through the short-to-long axis ratio of the  Tensor of Inertia \citep[ToI,][]{Metz07} concentration ellipsoid,  $c/a$ (and/or the r.m.s thickness normal to the plane, $\Delta$RMS, 
which are often correlated), provided there is a high intermediate-to-long axis ratio, $b/a$, which confirms a planar-like spatial distribution of satellites.
Quality comparisons 
between  planes
are done either considering constant $N_{\rm sat}$ (where a lower  $c/a$ means higher quality) or constant $c/a$ (where   more populated planes have higher quality). In this way, we are able to single out new high-quality planes.

This method has been applied  to the positional data of MW and M31 confirmed satellites
\citep{McConnachie12,Pawlowski13}.
Two  predominant, collimated  over-density regions show up in each of their respective 4-galaxy-normal density plots.
They  reveal that both satellite samples are highly structured in planar-like configurations.
However,  they show very different patterns:
while satellites in the MW form basically one main polar structure,
M31 satellites are spatially distributed along \textit{two} distinct collections of planes, inclined with respect to  the M31 disc and roughly perpendicular to each other.

We find planes of satellites with higher qualities than those previously reported  with a given $N_{\rm sat}$.
More specifically, 
we find a combination of 11 MW satellites that spatially describe a  plane with much higher quality
 than that defined 
by the 11 classical  (the most massive) satellites.
Similarly for M31, we present a combination of 14 satellites 
with much lower $c/a$   and $\Delta$RMS   values
 than those of the plane noted by \citet{Ibata13} and \citet{Conn13}  (Ibata-Conn-14 plane)
 with the PAndAS survey.

For the first time, the second-most predominant planar structure  (Peak 2) found in M31 has been studied in detail  (Peak 1 was studied in \citet{Pawlowski13}).
This peak points to a satellite planar configuration whose normal direction aligns with the line-of-sight between the Sun and  M31, and therefore is viewed nearly 
face-on (we recall that the planar configuration defined by M31 Peak 1 --containing the well-known GPoA with   $N_{\rm sat}$=19-- is viewed nearly edge-on). 
Our  analysis reveals a rich plane structure, with quality behaviour in terms of $c/a$ and $\Delta$RMS versus $N_{\rm sat}$ similar 
to that found around Peak 1, 
despite being more affected by the radial Sun - satellite distance uncertainties due to its orientation.
The satellites contributing to this
second
 planar configuration have  overall a different identity than  those contributing to Peak 1.

In particular, 
  both $c/a$ and $\Delta$RMS increase sharply for $N_{\rm sat}> 16$  satellites around Peak 1, and 
 for $N_{\rm sat}> 18$    around Peak 2. 
Therefore we state that the planes of satellites with precisely  these  $N_{\rm sat}$ values represent the 
 \textit{highest quality} planar structures for each peak, among the confirmed satellites within 300 kpc of M31.
As  evidence,
  the planes' normals stay stable up to $N_{\rm sat}=19$ for Peak 1, and $N_{\rm sat}=18$ for Peak 2, and then they change.
Interestingly, the plane from Peak 1 with $N_{\rm sat}$=16 
is more populated and thinner than the Ibata-Conn-14 plane.
Moreover, the plane with $N_{\rm sat}$=18 from Peak 2
 presents 
 very similar properties to the GPoA 
  but consists of an overall different satellite sample.
 This plane of satellites
had not been informed for so far.

Finally, 
the  richer plane structure in the MW and M31 we report in this work  was found because we allow the mass of   satellites to play no role in our search.
Indeed, through correlation tests we find 
that mass is not a satellite property that determines its 4-galaxy-normal contribution
to the main  over-density regions
 (i.e., its membership or not to the respective best-quality planes),
 either in the MW or M31 cases.
This is expected, given that globular clusters and streams have been found to align as well with the VPOS \citep{Pawlowski12}.

\begin{table*}
\centering
\small
\caption{   ToI fitting properties  of the observed planes of satellites in the MW (classical, VPOSall, VPOS-3) and M31 (Ibata-Conn-14, GPoA):
Galactic coordinates of the normal to the plane;
uncertainty in the normal direction;
distance to the MW/M31;
root-mean-square height of the plane;
short-to-long ellipsoid axis ratio;
intermediate-to-long ellipsoid axis ratio;
number of satellites included in the plane.
 Information for the classical plane of MW satellites has been taken from
  \citet{Metz07} and \citet{Pawlowski16}.
 The information for all rest of planes has been 
  extracted from 
 \citet{Pawlowski13} (see their Table 3), where distance uncertainties are considered by sampling the distances 1000 times with a Gaussian distribution around their most-likely distance.}

\begin{tabular}{l c c c c c}
\toprule
    &   \multicolumn{3}{c}{MW} & \multicolumn{2}{c}{M31}  \\
\midrule
Name  & classical &   VPOSall & VPOS-3 & Ibata-Conn-14   & GPoA  \\
\hline
$\vec{n}(l,b)$ [$^\circ$]  & (-22.7, -12.7)     &(-24.4,  -3.3)  &  (-10.5,   -2.8) & (26.2,  7.8) & (25.8, 7.6)   \\
$\Delta_{\rm sph} n$ [$^\circ$]  & --   &1.12& 0.43&  1.00&  0.79  \\
$D_{\rm MW}$ [kpc]   & 8.3   &7.9 $\pm$ 0.3 &10.4 $\pm$ 0.2   &    --   &  30.1 $\pm$ 8.8 \\
$D_{\rm M31}$ [kpc]  &  --   & 637.3 $\pm$ 13.0 &509.9 $\pm$ 10.2 &  4.1  $\pm$ 0.7  &  1.3 $\pm$ 0.6  \\
$\Delta$RMS height [kpc]&  19.6 & 29.3 $\pm$ 0.4 &19.9 $\pm$ 0.3 &    14.2  $\pm$ 0.2&         13.6 $\pm$ 0.2  \\
$c/a$             &   0.18      &0.301 $\pm$ 0.004 &0.209 $\pm$ 0.002&     0.125 $\pm$ 0.014 &        0.107 $\pm$ 0.005 \\
$b/a$           &       0.53    &0.576 $\pm$ 0.007& 0.536 $\pm$ 0.006  &  0.578 $\pm$ 0.084  &   0.615 $\pm$ 0.058  \\
$N_{\rm sat}$   &      11 & 27  &24 &  14    & 19  \\
\bottomrule
\end{tabular}
\label{table_obspl}
\end{table*}


\begin{table*}
\centering
\small
\caption{ 
Same as Table  \ref{table_obspl} for the high quality planes singled out in this work. 
 }

\begin{tabular}{l c  c c c c c}
\toprule
        &   \multicolumn{6}{c}{[Host-Peak-$N_{\rm sat}$]}  \\
\midrule
Name                 & MW-1-11 &  MW-1-14 &   M31-1-14 & M31-1-16 & M31-1-23 & M31-2-18   \\
\hline
$\vec{n}(l,b)$ [$^\circ$]  & (-9.6, 1.1)  & (-9.7, 1.1)   &  (25.9, 7.2)         & (25.3, 7.4)   &  (14.4, 5.0)     &  (-68.2, -23.6)   \\
$\Delta_{\rm sph} n$ [$^\circ$]  & 0.15  &  0.15     &0.45    &0.51   & 1.40    & 3.16   \\
$D_{\rm MW}$ [kpc]   &15.0$\pm$0.1   &14.6$\pm$1.0  &  36.2$\pm$4.9 &     40.9$\pm$5.8 & 185.7$\pm$15.0   &  741.6$\pm$6.5    \\
$D_{\rm M31}$ [kpc]   &  464.5$\pm$6.3  &  465.2$\pm$6.3   & 1.1$\pm$0.4  & 0.6$\pm$0.5    &  2.3$\pm$1.8     & 34.9$\pm$11.2      \\
$\Delta$RMS height [kpc]  &11.6$\pm$0.1  & 10.3$\pm$0.1    &  7.0$\pm$0.2   & 7.7$\pm$0.2   & 21.4$\pm$1.0    &21.0$\pm$4.2     \\
$c/a$             & 0.095$\pm$0.001 & 0.096$\pm$0.001    & 0.051$\pm$  0.002 &0.059$\pm$0.002  &0.189$\pm$0.009 &0.155$\pm$0.032  \\
$b/a$           & 0.595$\pm$0.004   &0.592$\pm$0.004  & 0.553$\pm$0.046   &0.564$\pm$0.046  &0.725$\pm$0.038 &0.734$\pm$0.006  \\
$N_{\rm sat}$   &      11 & 14  &14 & 16       & 23       & 18  \\
\bottomrule
\end{tabular}
\label{table_obsThisWork}
\end{table*}

\section*{Acknowledgements}
This work was supported through  MINECO/FEDER (Spain)  AYA2012-31101,  AYA2015-63810-P and MICIIN/FEDER (Spain) PGC2018-094975-C21 grants.
ISS acknowledges funding from the European Union's Horizon 2020 research and innovation programme 
under the Marie Sklodowska-Curie grant agreement No. 734374 (LACEGAL-RISE) for a secondment at the Astrophysics group of Univ. Andr\'es Bello (Santiago, Chile), and  from the Univ. Aut\'onoma de Madrid for a stay at the Leibniz Institut fur Astrophysik Potsdam (Germany). She thanks Dr. Patricia Tissera and Dr. Noam Libeskind for kindly hosting her.


\bibliographystyle{mn2e}
\bibliography{archive_planes}

\end{document}